\documentclass{andromedaone}       

\journal{BSM}
\vol{2021}
\jyear{Egypt}
\pages{Hurgada} 

\usepackage[T1]{fontenc}
\usepackage{multirow}
\usepackage{booktabs}
\usepackage{mathtools}
\usepackage{adjustbox}
\usepackage{array}

\newcommand{\eqal}[1]{\begin{align}#1\end{align}}

\begin{document}

\title{Searching for new physics through neutrino non-standard interactions}

\author{Yong Du\auno{1}}
\address{$^1$CAS Key Laboratory of Theoretical Physics, Institute of Theoretical Physics, Chinese Academy of Sciences, Beijing 100190, P. R. China}

\begin{abstract}
Due to the absence of any definite signals of new physics at colliders and from precision measurements, it has gradually become more and more popular in the community to utilize the effective field theory (EFT) framework in searching for new physics in a model-independent manner. In this letter, working in the EFT framework and focusing on neutrino non-standard interactions (NSIs), we report our most recent results on these NSIs from considering terrestrial neutrino oscillation experiments Daya Bay, Double Chooz, RENO, T2K and NO$\nu$A, and precision measurements of $N_{\rm eff}$ from Planck and CMB-S4.
\end{abstract}

\maketitle


\section{Introduction}
Though very successful and has been very precisely tested from various experiments, the Standard Model (SM) of particle physics can only be a low-energy effective theory as, for example, it cannot explain neutrino masses that is essential for neutrino oscillations, dark matter that makes up about 25\% of the energy budget of our Universe, as well as the baryon asymmetry of our Universe that is essential for understanding the existence of beings. For this reason, the SM has to be extended to enlarge its particle content to account for aforementioned phenomena. Tremendous models have been invented and intensively studied in the past decades such as SUSY, 2HDM, seesaw models etc., but unfortunately, all the experiments so far have reported a null result for any new physics, implying that if existing, the new physics could be too heavy to be directly produced from current experiments. As a consequence, EFTs could serve as an ideal framework in searching for new physics in a model-independent manner.

The SM EFT (SMEFT), obtained by integrating out the heavy newly introduced particles above the weak scale $\Lambda_{W}$, respects the SM gauge group $\rm SU(3)_c\times SU(2)_L\times U(1)_Y$. Besides the SM Lagrangian, the SMEFT also consists of a tower of higher dimensional operators that each respects the SM gauge group. The SMEFT can be parameterized as follows
\eqal{\mathcal{L}=\mathcal{L}_{\rm SM}+\sum\limits_{j\ge5}\frac{C_j}{\Lambda^{j-4}}\mathcal{O}^{(j)},\label{SMEFTeq}}
where $C_j$'s are the Wilson coefficients, $\Lambda$ is the characteristic scale of new physics and $j$ the dimension of operators. In the case of $j=5$, one has the well-known Weinberg operators\,\cite{Weinberg:1979sa} that are responsible for neutrino masses. Now at a much lower scale below $\Lambda_W$ and above 2\,GeV, one can further integrate out the top quark, the Higgs boson and the SU(2) gauge bosons in the SMEFT, resulting in the Low-energy EFT (LEFT) which now respects the $\rm SU(3)_c\times U(1)_{\rm EM}$ gauge group.

These higher dimensional operators in the LEFT introduce NSIs to the processes predicted in the SM. Among these operators, neutrino NSIs have gain significant attention in the past decades due to the observation of neutrino oscillations. In general, these neutrino NSIs can be classified into two categories: Charge-Current (CC) ones and Neutral-Current (NC) ones. On one hand, both CC and NC NSIs could modify neutrino oscillation probabilities and thus change the fitting of the oscillation parameters. These parameters are currently very precisely measured from neutrino oscillation experiments such as Daya Bay\,\cite{An:2012bu,An:2013zwz,Adey:2018zwh}, Double Chooz\,\cite{Ardellier:2006mn,DoubleChooz:2019qbj}, RENO\,\cite{Ahn:2012nd,Bak:2018ydk}, T2K\,\cite{Abe:2019vii,Abe:2011ks} and NO$\nu$A\,\cite{Ayres:2007tu,Acero:2019ksn} in the standard three active neutrino oscillation picture. Given these precision measurements, these NSIs, or equivalently the new physics, would have the chance to unveil themselves indirectly through causing some deviations from the fiducial SM prediction that can be measured from these neutrino oscillation experiments.

On the other hand, in the early Universe where the active particles are electrons, photons and neutrinos, NC NSIs between neutrinos and electrons/photons or NC NSIs among neutrinos of different flavors could also modify neutrino decoupling, which in turn would result in a different prediction of $N_{\rm eff}$, viz., the effective number of relativistic species in the early Universe. However, since $N_{\rm eff}$ is currently very stringently constrained by Planck\,\cite{Aghanim:2018eyx} and LEP\,\cite{ALEPH:2005ab} and would be measured at the percent level in the future by CMB-S4\,\cite{Abazajian:2016yjj}, one naturally expects that $N_{\rm eff}$ could also serve as an indirect probe in investigating new physics in the LEFT framework.

In this letter, based on our recent work\,\cite{Du:2020dwr,Du:2021idh}, we will discuss the sensitivity on new physics from both CC and NC neutrino NSIs. For the former, we will mainly focus on terrestrial neutrino oscillation experiments Daya Bay, Double Chooz, RENO, T2K and NO$\nu$A, and for the latter, we study their corrections to $N_{\rm eff}$ and take Planck and CMB-S4 into account for obtaining the constraints on new physics. We find that, for CC NSIs, terrestrial neutrino oscillation experiments are currently approaching new physics scale at $\sim$20\,TeV, while precision measurements of $N_{\rm eff}$ are now sensitive to new physics at the $\sim$200\,GeV scale.

\section{Parameterization of Neutrino non-standard interactions}

\subsection{CC NSIs}\label{subsec:CCNSIs}
Since neutrino oscillation experiments are performed at a very low energy scale, the LEFT, which is obtained by integrating out the heavy new particles in the UV theory, the top quark, the Higgs boson and the SU(2) gauge bosons, can be used for studying terrestrial neutrino oscillation experiments model-indenpendently. The Lagrangian relevant for neutrino oscillations from CC NSIs can be parameterized as\,\cite{Falkowski:2019kfn}
\begin{align}
\mathcal{L}_{\rm CC} \supset &-\frac{2 V_{u d}}{v^{2}}\left\{\left[\mathbf{1}+\epsilon_{L}\right]^{ij}_{\alpha \beta}\left(\bar{u}_i \gamma^{\mu} P_{L} d_j\right)\left(\bar{\ell}_{\alpha} \gamma_{\mu} P_{L} \nu_{\beta}\right)+\left[\epsilon_{R}\right]^{ij}_{\alpha \beta}\left(\bar{u}_i \gamma^{\mu} P_{R} d_j\right)\left(\bar{\ell}_{\alpha} \gamma_{\mu} P_{L} \nu_{\beta}\right)\right.\nonumber \\
& + \frac{1}{2}\left[\epsilon_{S}\right]^{ij}_{\alpha \beta}(\bar{u}_i d_j)\left(\bar{\ell}_{\alpha} P_{L} \nu_{\beta}\right)-\frac{1}{2}\left[\epsilon_{P}\right]^{ij}_{\alpha \beta}\left(\bar{u}_i \gamma_{5} d_j\right)\left(\bar{\ell}_{\alpha} P_{L} \nu_{\beta}\right)\nonumber \\
&+\left.\frac{1}{4}\left[\epsilon_{T}\right]^{ij}_{\alpha \beta}\left(\bar{u}_i \sigma^{\mu \nu} P_{L} d_j\right)\left(\bar{\ell}_{\alpha} \sigma_{\mu \nu} P_{L} \nu_{\beta}\right)+\mathrm{h.c.}\right\},\label{eq:CC}
\end{align}
where $V_{ud}$ is the ``ud'' component of the CKM matrix, $i,j = \{1,2,3\}$ the flavor of quarks, $\alpha, \beta =\{e,\mu,\tau\}$ that of charged leptons and neutrinos and $P=P_{L,R}$ are the chiral projection operators. The neutrino NSIs are parameterized by $\epsilon_{L,R,S,P,T}$, and to be generic, we also include interactions besides the $V-A$ type ones in the above Lagrangian.

\begin{table}[!t]
\tbl{Upper bounds of the NSI parameters associated with the neutrino production and detection. All constraints are given at 95\% CL.\label{NSIbounds}}{
\centering
\begin{tabular}{ccl}\hline\hline
NSI parameter & Upper bound & Experiments \\ \hline
$\left|\epsilon^s_{\mu e}\right|$ & 0.004 & \multirow{3}{*}{T2K\,\cite{Abe:2020vdv,Abe:2019vii,Zarnecki:2020yag}, NO$\nu$A\,\cite{Acero:2019ksn}} \\
$\left|\epsilon^s_{\mu \mu}\right|$ & 0.021 & {} \\
$\left|\epsilon^s_{\mu \tau}\right|$ & 0.080 &  {}\\
\hline\hline
$\left|\epsilon^d_{e e}\right|$ & 0.007 &  \\
$\left|\epsilon^d_{\mu e}\right|$ & 0.018 &  \\
$\left|\epsilon^d_{\tau e}\right|$ & 0.021 &  \multirow{1}{*}{Daya Bay\,\cite{An:2012bu,Adey:2018zwh}, Double Chooz\,\cite{Ardellier:2006mn,DoubleChooz:2019qbj},}\\
$\left|\epsilon^s_{ee}\right|$ & 0.007 &  \multirow{1}{*}{and RENO\,\cite{Ahn:2012nd,Bak:2018ydk}} \\
$\left|\epsilon^s_{e\mu}\right|$ & 0.018 &   \\
$\left|\epsilon^s_{e\tau}\right|$ & 0.021 & \\
\hline\hline
\end{tabular}}
\end{table}

To connect these CC NSIs to the neutrino oscillation probabilities that are formulated in the quantum mechanical framework in terms of the production parameter $\epsilon^s$ and the detection parameter $\epsilon^d$, one can follow the procedure developed in Ref.\,\cite{Falkowski:2019kfn} for a consistent matching between $\epsilon^{s,d}$ in the quantum mechanical formalism and the NSI parameters $\epsilon_{L,R,S,P,T}$ in the LEFT formalism. For reactor-type neutrino oscillation experiments like Daya Bay, Double Chooz and RENO and long-baseline neutrino oscillation experiments like T2K and NO$\nu$A, since the NSI parameters are very stringently constrained\,\cite{Du:2020dwr} as summarized in Table\,\ref{NSIbounds}, the linear order matching between these two formalisms would be enough for our purpose. To be specific, for the connection between $\epsilon^{s,d}$ and the CC NSI parameters $\epsilon_{L,R,S,P,T}$ for beta, inverse beta and pion decay, we cite the results in Ref.\,\cite{Falkowski:2019kfn} below and refer the readers to the Appendix of Ref.\,\cite{Du:2020dwr} for a detailed example for the calculation:
\begin{align}
\epsilon_{e \beta}^{s}=&\left(\epsilon_{L}-\epsilon_{R}-\frac{g_{T}}{g_{A}} \frac{m_{e}}{f_{T}\left(E_{\nu}\right)} \epsilon_{T}\right)_{e \beta}^{*},\quad{\text{~($\beta$ decay)}} \label{eq:sNSI1}\\
\epsilon_{\beta e}^{d}=&\left(\epsilon_{L}+\frac{1-3 g_{A}^{2}}{1+3 g_{A}^{2}} \epsilon_{R}-\frac{m_{e}}{E_{\nu}-\Delta}\left(\frac{g_{S}}{1+3 g_{A}^{2}} \epsilon_{S}-\frac{3 g_{A} g_{T}}{1+3 g_{A}^{2}} \epsilon_{T}\right)\right)_{e \beta},{\text{~(inverse $\beta$ decay)}}\label{inverseBetaDecayFor} \\
\epsilon_{\mu \beta}^{s}=&\left(\epsilon_{L}-\epsilon_{R}-\frac{m_{\pi}^{2}}{m_{\mu}\left(m_{u}+m_{d}\right)} \epsilon_{P}\right)_{\mu \beta}^{*},\quad{\text{~(pion decay)}}\label{eq:sNSI2}
\end{align}
with $g_S$, $g_A$ and $g_T$ the scalar, axial-vector and tensor charges of the nucleon, $\Delta\equiv m_n-m_p$ the difference between neutron and proton masses, $E_\nu$ the neutrino energy, and $f_{T}(E_\nu)$ the nucleon form factor resulting from the tensor-type NSI in eq.\,\eqref{eq:CC}.

\begin{figure}[!htb]
        \center{\includegraphics[width=0.6\textwidth]
        {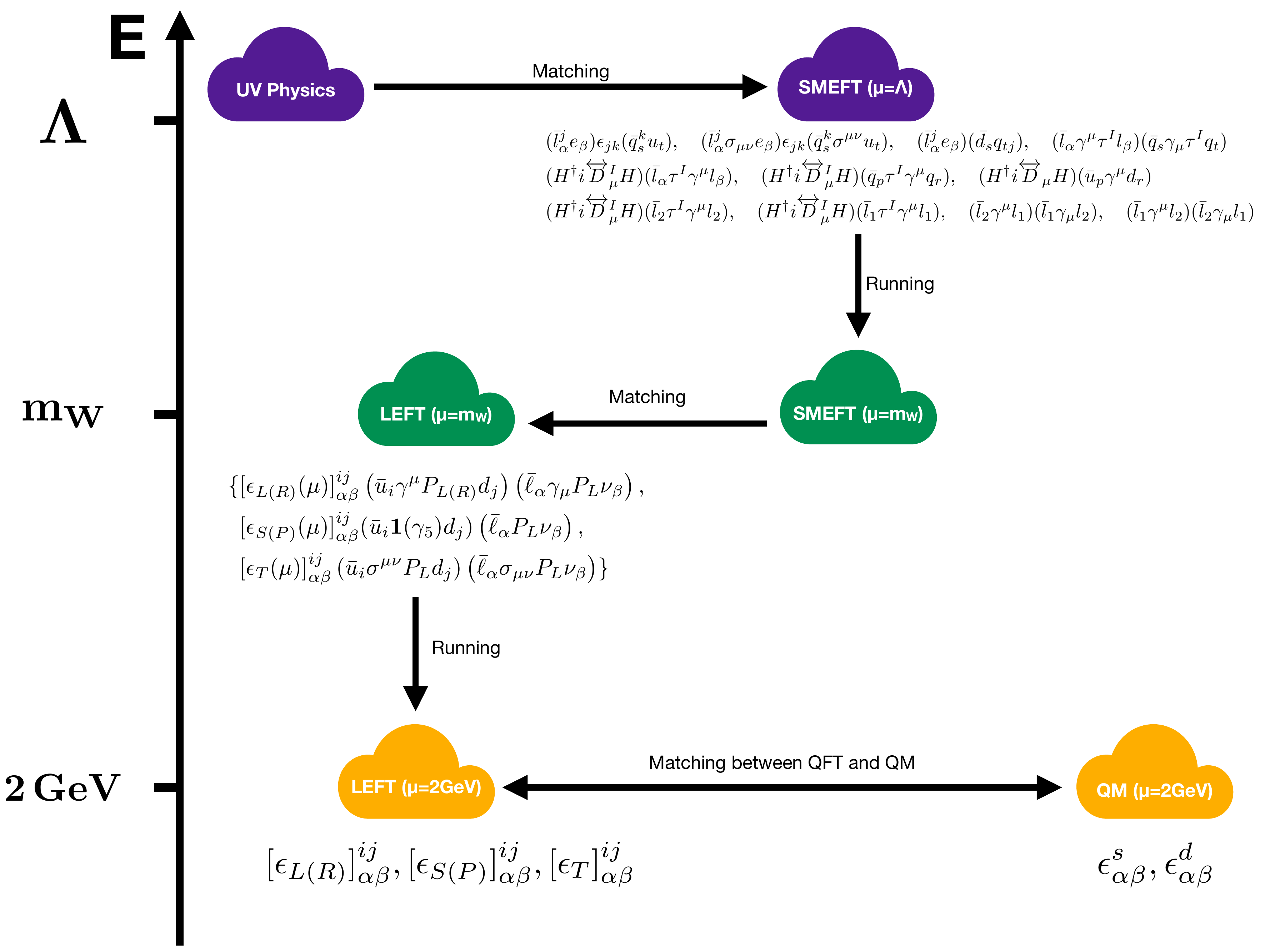}}
        \caption{A schematic description of how to relate the UV scale $\Lambda$ to the neutrino NSI parameters. The vertical axis defines the EFT scales. Figure adopted from Ref.\,\cite{Du:2020dwr}.}
        \label{fig:workflow}
  \end{figure}

With the matching formulae in eqs.\,(\ref{eq:sNSI1}-\ref{eq:sNSI2}), one can readily transfer the upper bounds on $\epsilon^{s,d}$ from neutrino oscillation experiments summarized in Table\,\ref{NSIbounds} onto these of $\epsilon_{L,R,S,P,T}$. Recall that $\epsilon_{L,R,S,P,T}$ are the LEFT Wilson coefficients obtained by integrating out the top quark, the Higgs boson and the SU(2) gauge bosons from the SMEFT, therefore, ultimately, $\epsilon_{L,R,S,P,T}$ are functions of the UV scale $\Lambda$ and the Wilson coefficients of the SMEFT in eq.\,\eqref{SMEFTeq}. Thus, from the upper bounds on $\epsilon_{L,R,S,P,T}$, one can obtain the lower bounds on the new physics scale $\Lambda$ once the Wilson coefficients are fixed. In addition, due to the large energy gap between the LEFT and the SMEFT, one needs to take the running effects into account to correctly gain any information on $\Lambda$. The running effects can be systematically included by solving the renormalization group equations. We show in Figure\,\ref{fig:workflow} the details of this procedure described above.

\subsection{NC NSIs}\label{subsec:ncnsi}
The effects of NC NSIs become significant for long-baseline neutrino oscillation experiments where matter effects start to impact. In literatures, the dimension-6 LEFT NC NSI operators are usually parameterized as follows:
\eqal{
\mathcal{L}_{\rm NSI}^{\rm NC}=-2\sqrt{2}G_F\sum_{\alpha,\beta,f,P}\epsilon_{\alpha\beta}^{f,P}\left(\bar{\nu}_\alpha\gamma_\mu P_L\nu_\beta\right)\left( \bar{f}\gamma^\mu P f \right)\label{lag:ncnsi},
}
with $f=e,u,d$ the charged fermions, $\alpha,\beta=e,\mu,\tau$ the flavor of neutrinos and $G_F$ the Fermi constant.

For a detailed study on the effects on neutrino oscillations from NC NSIs, see Ref.\,\cite{Du:2021rdg}. In this letter, as discussed in the introduction, we will study these NC NSIs from a different aspect by studying their effects on neutrino decoupling in the early Universe. To that end, we consider all possible LEFT NC NSI operators up to dimension-7. These operators are enumerated in Table\,\ref{SMEFTNeffOperators}.

\begin{table}
\tbl{LEFT operators relevant for $N_{\rm eff}$ up to dimension-7 with $\alpha,\beta,\alpha',\beta'=e,\mu,\tau$, the neutrino flavor indices, and $f=e$. The symbol ``$c$'' along with related operators means charge conjugation. Our convention for the Wilson coefficients are shown in the last column.\label{SMEFTNeffOperators}}{
 \renewcommand{\arraystretch}{1.5}
\centering
\begin{tabular}{|c|l|c|}
\hline
Dimensions & Operators & Wilson coefficients\\
\hline
{\rm dimension-5} & $\mathcal{O}_1^{(5)}=\frac{e}{8 \pi^{2}}\left(\bar{\nu}_{\beta} \sigma^{\mu \nu} P_{L} \nu_{\alpha}\right) F_{\mu \nu}$ & $C_1^{(5)}$ \\
\hline
\multirow{5}{*}{\rm dimension-6} & $\mathcal{O}_{1,f}^{(6)}=\left(\bar{\nu}_{\beta} \gamma_{\mu} P_{L} \nu_{\alpha}\right)\left(\bar{f} \gamma^{\mu} f\right)$ & $C_{1,f}^{(6)}$ \\
& $\mathcal{O}_{2,f}^{(6)}=\left(\bar{\nu}_{\beta} \gamma_{\mu} P_{L} \nu_{\alpha}\right)\left(\bar{f} \gamma^{\mu} \gamma_{5} f\right)$ & $C_{2,f}^{(6)}$ \\
& $\mathcal{O}_{3}^{(6)}=\left(\overline{\nu^c}_{\beta} P_{L} \nu_{\alpha}\right)\left(\overline{\nu^c}_{\beta'} P_{L} \nu_{\alpha'}\right)$ & $C_{3}^{(6)}$ \\
& $\mathcal{O}_{4}^{(6)}=\left(\bar{\nu}_{\beta} \gamma_{\mu} P_{L} \nu_{\alpha}\right)\left(\bar{\nu}_{\beta'} \gamma_{\mu} P_{L} \nu_{\alpha'}\right)$ & $C_{4}^{(6)}$\\
& $\mathcal{O}_{5}^{(6)}=\left(\overline{\nu^c}_{\beta} \sigma^{\mu \nu} P_{L} \nu_{\alpha}\right)\left(\overline{\nu^c}_{\beta'} \sigma^{\mu \nu} P_{L} \nu_{\alpha'}\right)$ & $C_{5}^{(6)}$\\
\hline
\multirow{9}{*}{\rm dimension-7} & $\mathcal{O}_{1}^{(7)}=\frac{\alpha}{12 \pi}\left(\overline{\nu^c}_{\beta} P_{L} \nu_{\alpha}\right) F^{\mu \nu} F_{\mu \nu}$ & $C_{1}^{(7)}$ \\
& $\mathcal{O}_{2}^{(7)}=\frac{\alpha}{8 \pi}\left(\overline{\nu^c}_{\beta} P_{L} \nu_{\alpha}\right) F^{\mu \nu} \widetilde{F}_{\mu \nu}$ & $C_{2}^{(7)}$ \\
& $\mathcal{O}_{5,f}^{(7)}=m_{f}\left(\overline{\nu^c}_{\beta} P_{L} \nu_{\alpha}\right)(\bar{f} f)$ & $C_{5,f}^{(7)}$ \\
& $\mathcal{O}_{6,f}^{(7)}=m_{f}\left(\overline{\nu^c}_{\beta} P_{L} \nu_{\alpha}\right)\left(\bar{f} i \gamma_{5} f\right)$ & $C_{6,f}^{(7)}$  \\
& $\mathcal{O}_{7,f}^{(7)}=m_{f}\left(\overline{\nu^c}_{\beta} \sigma^{\mu \nu} P_{L} \nu_{\alpha}\right)\left(\bar{f} \sigma_{\mu \nu} f\right)$  & $C_{7,f}^{(7)}$ \\
& $\mathcal{O}_{8,f}^{(7)}=\left(\overline{\nu^c}_{\beta} i\stackrel{\leftrightarrow}{\partial}_{\mu} P_{L} \nu_{\alpha}\right)\left(\bar{f} \gamma^{\mu} f\right)$ & $C_{8,f}^{(7)}$  \\
& $\mathcal{O}_{9,f}^{(7)}=\left(\overline{\nu^c}_{\beta} i\stackrel{\leftrightarrow}{\partial}_{\mu} P_{L} \nu_{\alpha}\right)\left(\bar{f} \gamma^{\mu}\gamma_5 f\right)$ & $C_{9,f}^{(7)}$  \\
& $\mathcal{O}_{10,f}^{(7)}=\partial_{\mu}\left(\overline{\nu^c}_{\beta} \sigma^{\mu \nu} P_{L} \nu_{\alpha}\right)\left(\bar{f} \gamma_{\nu} f\right)$ & $C_{10,f}^{(7)}$  \\
& $\mathcal{O}_{11,f}^{(7)}=\partial_{\mu}\left(\overline{\nu^c}_{\beta} \sigma^{\mu \nu} P_{L} \nu_{\alpha}\right)\left(\bar{f} \gamma_{\nu}\gamma_5 f\right)$ & $C_{11,f}^{(7)}$ \\
\hline
\end{tabular}}
\end{table}

Each operator in Table\,\ref{SMEFTNeffOperators} would modify the scattering and/or annihilation rates of neutrinos in the early Universe, thus also affect the energy injection rate (through neutrino-electromagnetic scattering) into the neutrino sector from the electromagnetic plasma or energy redistribution (through neutrino self-interactions) in the neutrino sector in the early Universe. As a consequence, the predicted effective number of relativistic species, $N_{\rm eff}$\cite{Shvartsman:1969mm,Steigman:1977kc,Mangano:2001iu} defined below, will also change:
\eqal{
\rho_R=\left[ 1+\frac{7}{8}\left(\frac{4}{11}\right)^{\frac{4}{3}} N_{\rm eff}\right]\rho_\gamma
}
where $\rho_\gamma$ is the photon energy density, and $\rho_R$ the total energy density from all relativistic species during the epoch of neutrino decoupling.

To obtain the exact prediction of $N_{\rm eff}$ with the inclusion of these operators in Table\,\ref{SMEFTNeffOperators}, one shall solve $\rho_R$ and $\rho_\gamma$ from the coupled Boltzmann equations, for which we follow Refs.\,\cite{Escudero:2018mvt,Escudero:2020dfa}. It turns out that the most challenging part in solving these Boltzmann equations comes from the collision term integrals that are extremely expensive to evaluate both numerically and analytically. For this reason and for the benefit of the community in the future, a complete, generic and analytic dictionary is provided in Ref.\,\cite{Du:2021idh}, which makes the Boltzmann equations simple ordinary differential equations that are numerically trivially solvable.

\section{Results}
\subsection{Constraints on dimension-6 SMEFT operators from CC NSIs}
\begin{figure}[t]
\centering{
\begin{tabular}{c}
\includegraphics[width = \textwidth]{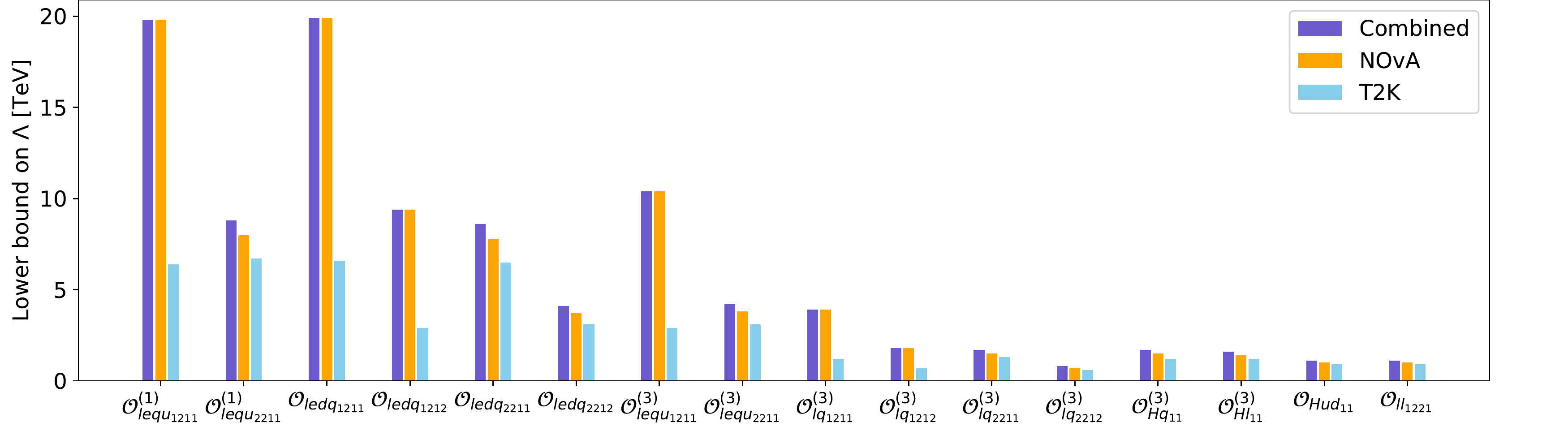}\\
\includegraphics[width = \textwidth]{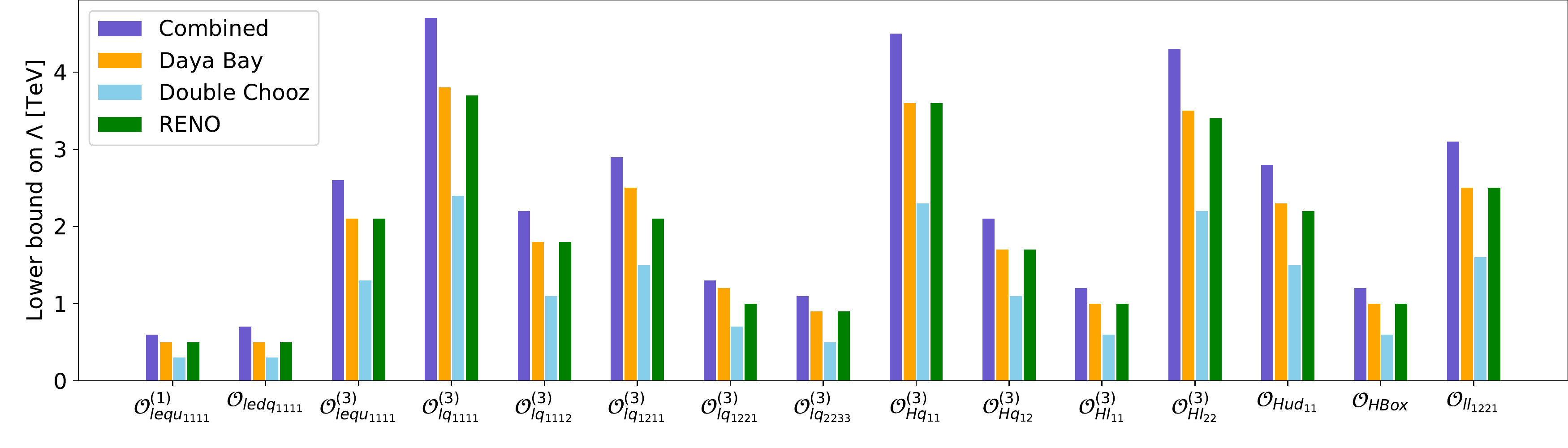}
\end{tabular}}
\caption{\label{fig:ccnsiconstr} Upper panel: Constraints on dimension-6 SMEFT operators in long-baseline neutrino oscillation experiments T2K and NO$\nu$A at 95\% CL. Lower panel: Constraints on dimension-6 SMEFT operators in reactor-type neutrino oscillation experiments Daya Bay, Double Chooz and RENO at 95\% CL. All constraints are obtained by fixing the Wilson coefficients at one. Plots adopted from Ref.\,\cite{Du:2020dwr}.}
\end{figure}

Following the procedure described in section\,\ref{subsec:CCNSIs} and depicted in Figure\,\ref{fig:workflow}, we study constrains on dimension-6 SMEFT operators from the CC NSIs by investigating their impacts on terrestrial neutrino oscillation experiments. We show our results on the UV scale $\Lambda$ in Figure\,\ref{fig:ccnsiconstr} by fixing the Wilson coefficients at unity and considering one non-vanishing SMEFT operator at a time. The upper panel is for long-baseline type neutrino oscillation experiments T2K and NO$\nu$A, and the lower one for reactor-type ones Daya Bay, Double Chooz and RENO.

From the upper panel, it is clear that currently the long-baseline type neutrino oscillation experiments are already exploring new physics around the 20\,TeV scale as indicated by the $\mathcal{O}_{lequ_{1211}}^{(1)}$ and $\mathcal{O}_{lequ_{2211}}^{(1)}$ operators. In contrast, as seen from the lower panel of Figure\,\ref{fig:ccnsiconstr}, the reactor-type ones are barely approaching the 5\,TeV scale. However, it is worth pointing out that these two types of experiments are sensitive to different subsets of the SMEFT operators. For illustration, while the reactor-type experiments constrain the $\mathcal{O}_{lq_{1111}}^{(3)}$ operator to be above about 5\,TeV, the long-baseline neutrino oscillation experiments are insensitive to this operator. We thus conclude that these two types of neutrino oscillation experiments are complementary to each other in searching for new physics.

\subsection{Constraints on NC NSIs from $N_{\rm eff}$}
\begin{figure}[t]
        \center{\includegraphics[width=0.7\textwidth]
        {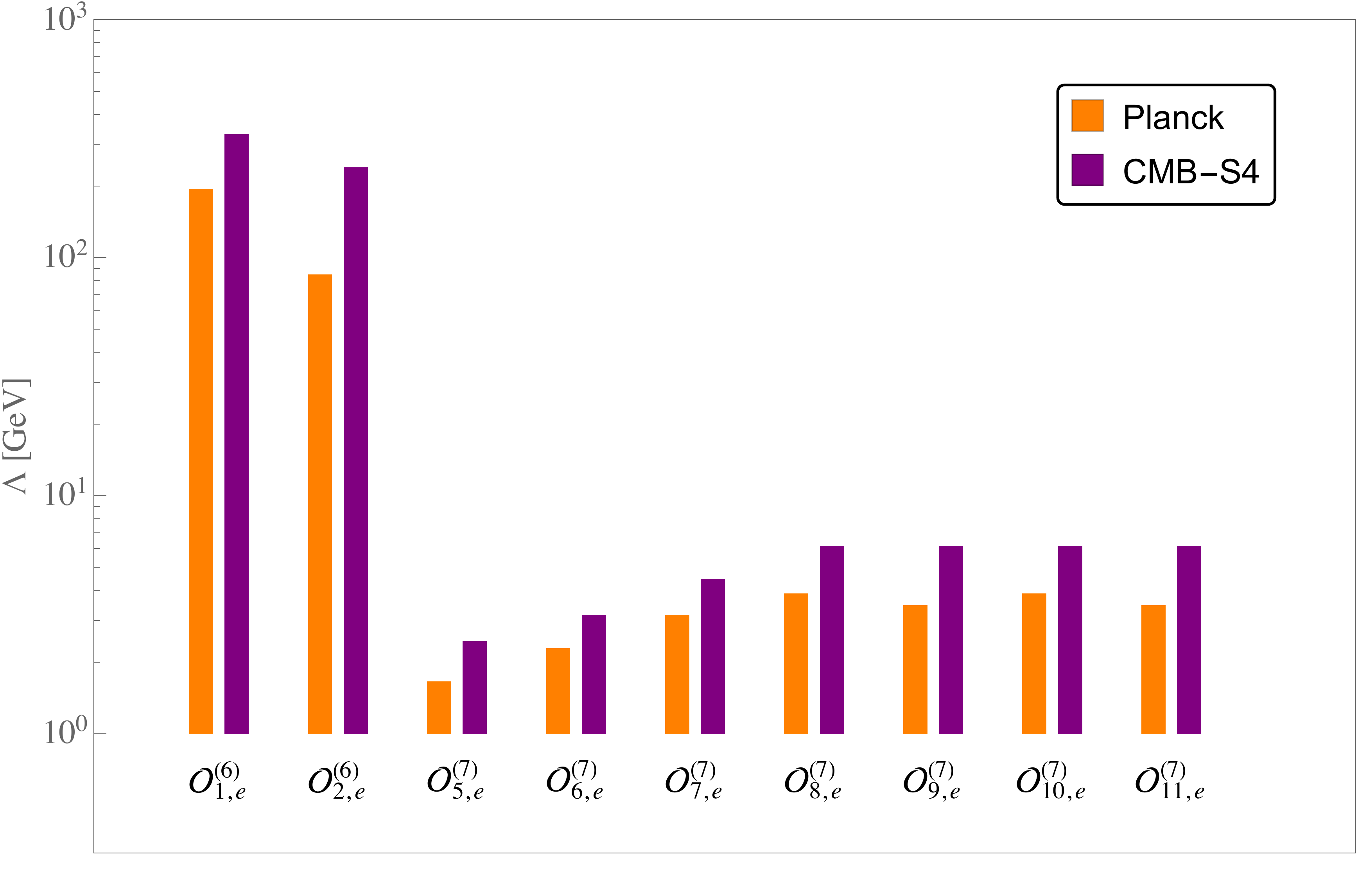}}
        \caption{Constraints on the UV scale $\Lambda$ from $\Delta N_{\rm eff}=N_{\rm eff}^{\rm SM+EFT}-N_{\rm eff}^{\rm SM}$, where $N_{\rm eff}^{\rm SM}$ is the SM prediction of $N_{\rm eff}$, and $N_{\rm eff}^{\rm SM+EFT}$ is that from SM and the LEFT operators in Table\,\ref{SMEFTNeffOperators}. All histograms here are obtained by fixing the Wilson coefficients at unity and considering only one non-vanishing operator only at a time. This plot is adopted from Ref.\,\cite{Du:2021idh}.}
        \label{plt:ConstraintsNP}
  \end{figure}

To obtain constraints on the NC NSIs, we first define the change in $N_{\rm eff}$ as
\eqal{
\Delta N_{\rm eff} = N_{\rm eff}^{\rm SM+EFT} - N_{\rm eff}^{\rm SM},
} 
where $N_{\rm eff}^{\rm SM+EFT}$ is the theoretical prediction of $N_{\rm eff}$ from the LEFT with the NC NSI operators in Table\,\ref{SMEFTNeffOperators}, and $N_{\rm eff}^{\rm SM}=3.044$\,\cite{Akita:2020szl,Froustey:2020mcq} is that from the SM. For Planck, we use the current result $N_{\rm eff}=2.99^{+0.34}_{-0.33}$\,\cite{Aghanim:2018eyx} at the 95\% CL, and $\Delta N_{\rm eff}<0.06$ at 95\% CL for CMB-S4\,\cite{Abazajian:2016yjj,Abazajian:2019tiv,Abitbol:2017nao,Abazajian:2019eic}. As discussed in section\,\ref{subsec:ncnsi}, after solving the Boltzmann equations with the help of the complete dictionary presented in Ref.\,\cite{Du:2021idh} one can readily obtain the corrections to $N_{\rm eff}$ from these LEFT NC NSI operators listed in Table\,\ref{SMEFTNeffOperators}. We show our results in Figure\,\ref{plt:ConstraintsNP} by requiring the change in $N_{\rm eff}$ to be within the uncertainties from current Planck data and the planned precision goal of CMB-S4. The results are shown in orange for Planck and purple for CMB-S4 in Figure\,\ref{plt:ConstraintsNP}.

From Figure\,\ref{plt:ConstraintsNP}, one concludes that the dimension-6 $\nu-e$ contact interacting operators $\mathcal{O}_{1,e}^{(6)}$ and $\mathcal{O}_{2,e}^{(6)}$ are relatively more constrained to be above $\sim$200\,GeV and $\sim$86\,GeV respectively. In addition, the constraint on the vector-type operator $\mathcal{O}_{1,e}^{(6)}$ is stronger than that on the axial-vector operator $\mathcal{O}_{2,e}^{(6)}$ since the former interfere with the SM constructively while the latter destructively: In the constructive case, neutrinos and electrons are effectively more tightly coupled such that the neutrinos would decouple earlier from the electromagnetic plasma than that in the destructive case, thus the neutrinos would be hotter at the time of decoupling and therefore a larger prediction of $N_{\rm eff}^{\rm SM+EFT}$. This explains why the lower bound on $\Lambda$ from the $\mathcal{O}_{1,e}^{(6)}$ operator is larger compared to that from the $\mathcal{O}_{2,e}^{(6)}$ operator.

The dimension-7 operators in Figure\,\ref{plt:ConstraintsNP} are only constrained to be below $\sim$10\,GeV from the current results of Planck as well as the planned precision goal of CMB-S4. The reason is that these operators are suppressed by one more power of $\Lambda$ compared with the dimension-6 ones. Nonetheless, these constraints on the dimension-7 operators shown in Figure\,\ref{plt:ConstraintsNP} are firstly obtained in Ref.\,\cite{Du:2021idh} and would serve as the starting point in the future when much higher precision measurements become avaliable.

Last but not least, the constraints on $\mathcal{O}_{8,e}^{(7)}$ and $\mathcal{O}_{10,e}^{(7)}$ turn out to be the same as can also be seen from Figure\,\ref{plt:ConstraintsNP}. This can be understood from their analytical expression for these two operators:
\eqal{\rho_{\nu}^{\rm tot.}(\mathcal{O}_{8,e}^{(7)})-\rho_{\nu}^{\rm tot.}(\mathcal{O}_{10,e}^{(7)})=\frac{960}{\pi^5\Lambda^6}\left( C_{8,e}^{(7)} - C_{10,e}^{(7)}\right)\left( C_{8,e}^{(7)} + C_{10,e}^{(7)}\right)\times \mathcal{F}(T_\gamma,T_{\nu_e},T_{\nu_\mu},\mu_{\nu_e},\mu_{\nu_\mu}),
}
where we omit the irrelevant tiny neutrino masses here and $\mathcal{F}(T_\gamma,T_{\nu_e},T_{\nu_\mu},\mu_{\nu_e},\mu_{\nu_\mu})$ is a function that only depends on photon temperature, neutrino temperatures and neutrino chemical potentials.\footnote{The explicit expression of $\mathcal{F}$ is lengthy and does not matter here. For its full expression, see the supplementary material of Ref.\,\cite{Du:2021idh}. It can be shown that these two operators are related by the equation of motion for the neutrinos as shown in Ref.\,\cite{Li:2021phq}, we thank Xiao-Dong Ma for pointing this out to us.} Clearly, these two operators contributes exactly the same to the total energy density of neutrinos when their Wilson coefficients are fixed to be the same. We point out that a similar observation also holds for the $\mathcal{O}_{9,e}^{(7)}$ and the $\mathcal{O}_{11,e}^{(7)}$ operators.

\section{Conclusions and outlook}
We investigate both CC and NC neutrino NSIs in this letter in the EFT framework. For CC NSIs, we study the effects of dimension-6 SMEFT operators on terrestrial neutrino oscillation experiments. By fixing the Wilson coefficients at unity and assuming only one operator dominate at a time, we find that long-baseline neutrino oscillation experiments T2K and NO$\nu$A are currently approaching new physics at the 20\,TeV scale while reactor neutrino oscillation experiments Daya Bay, Double Chooz and RENO are sensitive to new physics at $\sim$5\,TeV scale. Even though these two types of neutrino oscillation experiments are sensitive to new physics at much different scales, we stress that they are complementary to each other in the sense that they are sensitive to different subsets of SMEFT operators as indicated in Figure\,\ref{fig:ccnsiconstr}.

For NC NSIs, we study in detail their rule in neutrino decoupling in the early Universe for LEFT operators up to dimension-7. We find that the dimension-6 operators are more constrained compared with the dimension-7 ones as they are less suppressed by the UV scale $\Lambda$. Quantitatively, from the $\mathcal{O}_{1,e}^{(6)}$ operator, $\Lambda$ is constrained to be above about 200\,GeV while that from $\mathcal{O}_{2,e}^{(6)}$ is about 86\,GeV. The difference in these two operators comes from the fact the former interfere with the SM constructively while the latter destructively. Furthermore, constraints on dimension-7 operators as shown in Figure\,\ref{plt:ConstraintsNP} are one order of magnitude smaller than those on the dimension-6 operators, but these results are firstly obtained from our work in Ref.\,\cite{Du:2021idh} and would serve as the starting point in the future when higher precision measurements become possible.

We comment on that, in the future when long-baseline experiments like JUNO and DUNE start operation, the matter effects from these NC NSIs would start to impact. In this case, to obtain the correct bounds on the UV physics in the EFT framework, one has to include the matter effects from these NC NSIs. For a detailed study on this topic, see Ref.\,\cite{Du:2021rdg}. We also point out that the correlation among multiple operators may change the constraints on the UV scale $\Lambda$ by orders of magnitude\,\cite{Du:2020dwr}. In both the CC and NC NSI cases, it is worth obtaining a global fitting on multiple operators to fully understand the correlation among multiple operators and their impact on constraining the UV physics.

\section*{Acknowledgements}
The author is indebted to Hao-Lin Li, Jian Tang, Sampsa Vihonen and Jiang-Hao Yu for the collaboration. The author also wishes to thank the committee of the ``Beyond Standard Model: From Theory to Experiment (BSM-2021)'' workshop. This work is supported by the National Natural Science Foundation of China (NSFC) under Grants No.\ 12022514 and No.\ 11875003 and CAS Project for Young Scientists in Basic Research YSBR-006, by the National Natural Science Foundation of China (NSFC) under Grant No.\ 12047503, by the National Key Research and Development Program of China under Grant No. 2020YFC2201501, and also by the internal funds for postdocs from the CAS Key Laboratory of Theoretical Physics, Institute of Theoretical Physics, Chinese Academy of Sciences.

\bibliographystyle{unsrt}



\end{document}